\documentclass{elsart}

\usepackage{natbib}
\usepackage{graphicx}
\usepackage{amssymb}

\journal{New Astronomy}

\begin{document}

\def\lsim{\lower.5ex\hbox{$\; \buildrel < \over \sim \;$}}
\def\gsim{\lower.5ex\hbox{$\; \buildrel > \over \sim \;$}}
\def \simeq{\lower.3ex\hbox{$\; \buildrel \sim \over - \;$}}
\def\eg{{\it e.g.,} }
\def\etal{{\em et al.} }
\def\ie{{\em i.e.,} }
\def\xci{{$x_{ci}~~$}}
\def\xco{{$x_{co}$}}

\begin{frontmatter}

\title{Computation of mass loss from viscous accretion disc in presence of cooling}

\author{Santabrata Das}
\address{ARCSEC, Sejong University, Seoul, South Korea}
\ead{sbdas@canopus.cnu.ac.kr}

\author{Indranil Chattopadhyay \corauthref{cor}}
\address{Department of Astronomy and Space Science, Chungnam National Univ,
Daejeon, South Korea}
\ead{indra@canopus.cnu.ac.kr}
\corauth[cor]{Corresponding author.}

\begin{abstract}
Rotating accretion flow may undergo centrifugal pressure
mediated shock transition even in presence of various
dissipative processes, such as viscosity and
cooling mechanism. 
The extra thermal gradient force along the vertical direction
in the post shock flow drives a part of the accreting matter as bipolar
outflows which are believed to be the precursor of relativistic jets.
We compute mass loss rates from a viscous accretion disc in presence of
synchrotron cooling in terms of the inflow parameters. We show
cooling significantly affects the mass outflow rate, to the extent that,
jets may be generated from flows with higher viscosity.
We discuss that our formalism may be employed to explain
observed jet power for a couple of black hole candidates.
We also indicate that using our formalism,
it is possible to connect the spectral properties of the disc with
the rate of mass loss.
\end{abstract}


\end{frontmatter}

\section{Introduction}

In recent years, it has been established that AGNs and Microquasars suffer
mass loss in the form of jets and outflows \citep{f98,mr99}.
Generation of jets or outflows around gravitating centres with hard boundaries
(\eg neutron stars, YSOs etc.)
are quite natural, however, it is altogether a different
proposition to consider the same
around a black hole. As black holes do not have either hard boundaries or
intrinsic atmospheres, jets/outflows have to originate from the accreting
matter onto black holes, though there is no consensus
about the exact mechanism of jet formation.
One of the motivation of studying black hole accretion is therefore to
understand the primary mechanism in the accretion process
which may be responsible for
the generation of jets. 
In addition, recent observations have established that,
whatever be the exact mechanism behind the formation of jets/outflows
around black holes, the formation of jets is intrinsically linked with
spectral states of the associated black hole candidates.
In particular, \citet{gfp03} showed that quasi steady jets are generally ejected
in the hard state, which suggests
that the generation or quenching of jets do depend on 
various states of the accretion disc.
Several theoretical attempts were made to explain the possible
mechanisms of jet generation
from accretion disc.
\citet{xc97} reported the formation of outflows by considering self-similar
solutions. 
\citet{c99,dc99} estimated mass outflow rates in terms of
inflow parameters from an inviscid advective
disc. In particular, these authors showed that
the centrifugal barrier may produce shock, and the post-shock disc
can generate bipolar outflows. They also showed
mass outflow rates depend on the strength of the centrifugal
barrier, as well as, its thermal driving.
\citet{dcnc01} extended this work
to show that such outflows generated by accretion shock is compatible
with the spectral state of the accretion disc. 
The shock induced relativistic outflows could be obtained if
various acceleration mechanism,
namely, first order Fermi acceleration at the shock \citep{lb05},
or radiation pressure \citep{c05}, are considered.

Recently, \citet{cd07} computed mass outflow rates
from a viscous advective disc and showed that
the mass outflow rate decreases with the increase of viscosity
parameter.
In realistic accretion disc, a variety of dissipative processes
are expected to be present, and viscosity is just one of them.
In absence of mass loss, \citet{gl04} conjectured that
cooling processes will not affect the nature of 
advective accretion solutions. However, \citet{d07} explicitly
showed that cooling processes play a crucial role in determining
the flow variables as well as the shock properties.
Therefore, it will be worthwhile to investigate, how cooling would affect
the mass outflow rate from a viscous accretion disc.
In presence of viscosity, as matter flows inward angular momentum decreases
while specific energy increases. A cooling process unlike viscosity,
only reduces the energy of the flow
and leaves the angular momentum distribution un-affected. 
Thus the increase of flow energy due to viscous heating may be abated
by incorporating cooling mechanism.  
As cooling is more efficient at the hotter and denser
post-shock region (abbreviated as CENBOL $\equiv$
CENtrifugal pressure supported BOundary Layer),
the decrease of CENBOL energy will be more pronounced compared to
the pre-shock energy.
In reality, more energetic flows at the outer edge,
which do not
satisfy shock conditions in absence of cooling,
may undergo shock transition in its presence.
Consequently, 
more energetic CENBOL may be produced for flows with higher cooling efficiency,
and hence there is a possibility of
enhanced jet driving.
In this paper, we would like to address
these issues in detail.

In the next section, we present the model assumptions and the governing
equations. In Section 3, we discuss the methodology of computing
self-consistent inflow-outflow solutions and
present the solutions. In Section 4, we apply our formalism on two
black hole candidates to compute the mass outflow rate,
and compare it with the observed jet power.
In the last section we draw concluding remarks.

\section{Model Assumptions and Equations of motion}

In a disc-jet system, there are two separate flow geometries, namely,
one for accretion flows and the other for outflows. Axis-symmetry
and steady state conditions are assumed for the disc-jet system.
In the present paper, we consider thin, viscous accretion flow
in presence of synchrotron cooling. Jets are assumed to be tenuous.
Since jets are in general collimated, they should have less angular momentum
and therefore less differential rotation compared to the accretion
disc. Thus, we ignore the effect of viscosity in jets.
As jets are believed to originate from the inner part of the disc,
which in our model is the CENBOL, the jet base must be described by identical local
accretion flow variables (see section 3), \ie the specific energy, the angular momentum etc of the CENBOL.
Consequently, we neglect the torque between the disc and the jet at the
jet base. 
It is to be remembered that, to keep the jets collimated, angular momentum will be reduced either by magnetic field (stochastic fields, considered in the paper, are not effective in doing so), or by radiation [see, \citep{c05}], however these processes
have not been considered here.
In reality, back reactions on the disc in the form of extra
torque at the jet base and/or feedback effect from failed jets
are not altogether ruled out. To study these effects, one requires to undertake
numerical simulation, which is beyond the scope of the present frame work.
Moreover, jets are supposed to be colder than the accretion discs.
Therefore, we assume jets to be adiabatic, at least up to its critical
point. We use pseudo-Newtonian potential introduced by \citet{pw80}
to approximate the space time geometry around a non-rotating black hole.

\begin{figure}

\begin{center}
\includegraphics[width=0.58\textwidth]{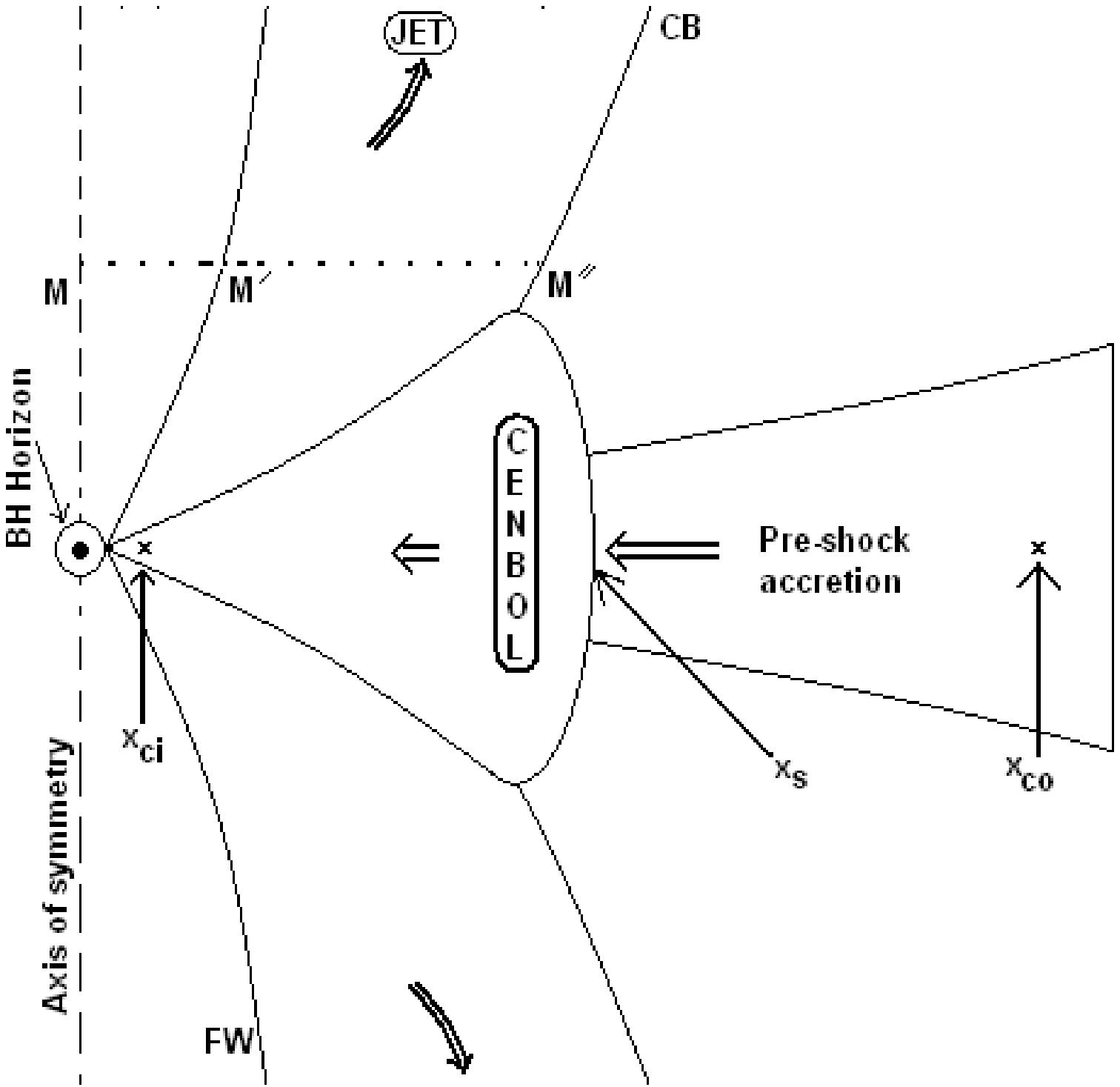}
\end{center}

\caption{ A schematic diagram of disc-jet system. The outer
and inner critical points
$x_{co}$ and $x_{ci}$ are marked in the figure. The shock is located at
$x_s$. The jet geometry is bounded by FW and CB. MM$^{\prime}=x_{FW}$
and MM$^{\prime \prime}=x_{CB}$ (described in the text).}
\label{fig1}
\end{figure}

A schematic structure of shocked advective accretion disc and the associated
jet are presented in Fig. 1. Here, $x_{co}$ and $x_{ci}$
are the outer and the inner critical points of the disc, respectively.
The centrifugal pressure acts as a `barrier' to
the supersonic matter at $x_{ci}<x<x_{co}$ and a shock at $x_s$ is formed.
The post-shock disc
is indicated in the figure as CENBOL. At the shock, matter momentarily
slows down and
ultimately dives into the black
hole supersonically through $x_{ci}$. Excess
thermal driving in CENBOL drives a fraction of accreting matter as bipolar
jet which flows
within two geometric surfaces called the Funnel Wall (FW) and
the Centrifugal Barrier
(CB) \citep{mlc94,mrc96}.

The system of units used in this paper is $2G=M_{\rm BH}=c=1$,
where $G$, $M_{\rm BH}$ and $c$ are the universal gravitational
constant, the mass of the black hole and the speed of light, respectively.
Since we use the geometrical system of units, our formalism is applicable
for both the galactic and the extra galactic
black hole candidates.
Two separate sets of hydrodynamic equations for accretion and
jet, are presented bellow.

The dimensionless hydrodynamic equations
that govern the motion of accreting matter are \citep{gut,d07},

\noindent the radial momentum equation :
$$
u \frac {du}{dx}+\frac {1}{\rho}\frac {dP}{dx}
-\frac {\lambda^2(x)}{x^3}+\frac {1}{2(x-1)^2}=0,
\eqno(1a)
$$
where, $u$, $\rho$, $P$, and $\lambda(x)$ are the radial
flow velocity, the local density, the isotropic pressure and
the local specific angular momentum, respectively. Here $x$
is the cylindrical radial coordinate.

\noindent The baryon number conservation equation :
$$
{\dot M} = 2 \pi {\Sigma} u x,
\eqno(1b)
$$
where, ${\dot M}$ and $\Sigma$ are the mass accretion rate
and the vertically integrated density, respectively.
In our model, the accretion rates in the pre shock and post shock regions
are different as some fraction of the accreting matter is ejected as
outflow. Actually, the post-shock matter is flown into two channels ---
one is the accreting part (falling onto black holes through $x_{ci}$)
and the other is the outflowing part \citep{mlc94,mrc96,cd07}.
More specifically, the combination of accretion and outflow rate in
the post shock region remain conserved with the pre-shock accretion
rate (see Eq. 3). 

\noindent The angular momentum conservation equation :
$$
u \frac {d\lambda(x)}{dx}+\frac{1}{\Sigma x}
\frac {d}{dx}\left( x^2 W_{x\phi}\right)=0,
\eqno(1c)
$$
where, $W_{x\phi}(=- \alpha \Pi)$ denotes the viscous stress,
$\alpha$ is the viscosity parameter and $\Pi$ is the vertically
integrated total (\ie thermal$+$ram) pressure. The viscosity
prescription employed in this paper was developed by \citet{cm95}
and has been employed to study advective accretion disc by a group
of workers \citep{gut,cd04,gl04,d07,cd07}. This viscosity
prescription is more suitable for flows with significant
radial velocity as it maintains angular momentum distribution
continuous across the shock unlike Sakura-Sunyaev type viscosity
prescription which was proposed for a Keplerian disc.

\noindent And finally,
the entropy generation equation :
$$
u T \frac {ds}{dx}= Q^+ - Q^-,
\eqno(1d)
$$
where, $s$ is the specific entropy of the flow, $T$ is the local temperature.
$Q^+$ and $Q^-$ are the heat gained and lost by the flow, and are given
by \citep{gut,d07,st83},
$$
Q^+=-\frac{{\alpha}}{\gamma}
x(ga^2+\gamma u^2)\frac{d\Omega}{dx}
$$
and
$$
Q^-=\frac{\beta S_ia^5}{ux^{3/2}(x-1)}.
$$
Here, $g=I_{n+1}/I_n$, $n=1/(\gamma -1)$, $I_n=(2^nn!)^2/(2n+1)!$
\citep{matetal84}, and $\gamma$($=4/3$) is the adiabatic index.
Presently, we consider only synchrotron cooling.
In the above equation, $\beta$ is the cooling parameter, and
$S_i$ is the synchrotron cooling term which is independent of the
flow variables and is given by,
$$
S_i=\frac{32 \eta {\dot m}_{i}{\mu}^2e^41.44{\times}10^{17}}
{3{\sqrt{2}}m_e^3{\gamma}^{5/2}}
\frac{1}{2GM_{\odot}c^3},
$$
where, $e$ is the electron charge, $m_e$ is electron mass,
${\dot m}_{i}$ is the accretion rate in units of Eddington rate, 
$M_{\odot}$ is solar mass, and for fully ionized plasma $\mu =0.5$.
The suffix `$i=\mp$' represents the quantities in the pre/post shock
disc region. It is to be borne in mind that in absence of shock
${\dot m}_+={\dot m}_-$, therefore $S_+=S_-$.
Due to the uncertainties of the realistic magnetic field
structure in the accretion disc, we have assumed stochastic magnetic field.
The ratio between the magnetic pressure and the gas pressure is represented
by $\eta$. 
The magnetic field strength is estimated by assuming partial
equipartition ($\eta{\leq}1$)
of the magnetic pressure with the gas pressure. 
In this paper, we have ignored bremsstrahlung cooling, since it is
a very inefficient cooling process \citep{cc00,dc04}.
The expression for bremsstrahlung cooling \citep{rl79} in vertical
equilibrium is given by,
$$
Q^-_B=\frac{B_i}{ux^{3/2}(x-1)},
$$
where
$$
B_i=\frac{2.016{\times}10^{-10}}{4{\pi}m^2_p}
(\frac{\mu m_p}{2k_{\small B}})^{1/2}\frac{{\dot m}_{i}}{2GM_{\odot}c},
$$
where, $m_p$ is the proton mass and $k_{\small B}$ is the Boltzmann
constant.
For identical accretion rates
$$
\frac{S_i}{B_i}=3.26{\times}10^7{\times}{\eta}.
$$
Therefore, it is quite evident that the synchrotron cooling is much
stronger than bremsstrahlung. However, bremsstrahlung photons may
interact with the accreting gas itself and in that sense bremsstrahlung
may be important. Such complicated situation is not addressed in the
present paper. We have also not considered inverse-Compton, since that
will require a proper two temperature solution which is also beyond the
scope of the present effort.

In the present paper, we have chosen ${\dot m}_{-}=0.1$ and
$\eta=0.1$ as the representative case, until stated otherwise.

Under the adiabatic assumption for the jet,
the momentum balance equation can be represented in the following
integrated form: 
$$
{\mathcal E}_j= \frac{1}{2}v^2_j+na^2_j +\frac {\lambda^2_j}{2x^2_j}
-\frac{1}{2(r_j-1)},
\eqno{(2a)}
$$
where, ${\mathcal E}_j$ and $\lambda_j$ are the specific energy and
angular momentum
of the jet, respectively. Other flow variables are the jet velocity ($v_j$)
and sound speed ($a_j$).
Furthermore, $x_j[=(x_{CB}+x_{FW})/2]$ and
$r_j[=(x^2_j+y^2_{CB})^{1/2}]$ are the cylindrical and spherical
radius of the jet streamline. 
The functional form of the coordinates of CB and FW are 
[see, \citet{cd07}],
$$
x_{CB}=\left[ 2 \lambda^2_j r_{CB} (r_{CB}-1)\right]^{1/4},
$$
$$
x^2_{FW}=\lambda^2_j \frac{(\lambda^2_j-2)+
\sqrt{(\lambda^2_j-2)^2-4(1-y^2_{CB})}}{2},
$$
where, $x_{CB}$ and $x_{FW}$ are measured at the same height 
of jet streamline and is given by
$y_{CB}=\sqrt{(r^2_{CB}-x^2_{CB})}$.

The integrated form of mass-flux conservation equation for the
jet is given by,
$$
{\dot M}_{\rm out}=\rho_j v_j {\mathcal A},
\eqno{(2b)}
$$
where, ${\dot M}_{\rm out}$ is jet outflow rate and $\rho_j$ is the 
local density of the jet.
The jet cross-sectional area is given by,
${\mathcal A}=2\pi (x^2_{CB}-x^2_{FW})$.

\section{Accretion-Ejection solution}

It is well known that matter falling onto black holes have to cross
one or more critical points depending on the absence or presence
of shock transition \citep{gut,cd04,cd07}. If the flow parameters
allow shock transition then matter must cross the sonic horizon
twice, once before the shock and then after the shock. The location
of the latter is called the inner critical point ($x_{ci}$) and
the former is known as outer critical point ($x_{co}$). In absence
of dissipation, the energy (${\mathcal E}$) and angular momentum
($\lambda$) of the flow is conserved, and therefore $x_{ci}$ and/or
$x_{co}$ are uniquely obtained in terms of ${\mathcal E}$ and
$\lambda$, and consequently all possible flow solutions. ${\mathcal E}$
and $\lambda$ do not remain conserved along a dissipative flow and
therefore critical points cannot be determined uniquely. To obtain
solutions of a dissipative accretion flow in a simpler way, one needs
to know at least one set of critical point parameters (\eg$x_c$,
$\lambda_c$). Fortunately, the range of ($x_{ci}$,$\lambda_{ci}$)s
varies from ($2r_g\lsim x_{ci}$ $\lsim4r_g$,$1.5\lsim
{\lambda_{ci}}\lsim \lambda_{ms}$), where $\lambda_{ci}$,
$\lambda_{ms}$ are the angular momentum at the inner critical point
and the marginally stable orbit, respectively [\eg \citet{c89,gut,cd04}].
Here $r_g$ is the Schwarzschild radius. Therefore for a viscous flow,
it is easier to consider $x_{ci}$ and $\lambda_{ci}$ as parameters
for solving the flow equations, along with the viscosity parameter
${\alpha}$ \citep{cd04,cd07}. In presence of cooling,
one should also supply the accretion rate at $x_{ci}$ in addition to
($x_{ci}, \lambda_{ci}, \alpha$). Presently, we fix
accretion rate and vary $\beta$ to study the effect of cooling.
Hence the existence of $x_{co}$ can be obtained only in presence
of a shock. 

In this paper, we consider infinitesimally thin adiabatic shock,
generally expressed by the continuity of energy flux, mass flux
and momentum flux across the shock, and is generally called Rankine-Hugoniot
(RH) shock conditions. Numerical simulations [\eg \citet{eetal85,
mlc94, mrc96}] have shown that thermally driven outflows could originate
from the hot inner part of the disc.
When rotating matter accretes towards the black hole,
centrifugal force acts as a
barrier, inducing the formation of shock. At the shock, flow temperature rises
sharply as the kinetic energy of the flow is converted into the thermal
energy. This excess thermal energy may drive a significant fraction of
accreted material as outflows. 
Thus bulk properties such as excess thermal driving along $z$ direction
is a legitimate process for mass ejections.

The modified Rankine-Hugoniot shock conditions in presence of mass loss
are [\citet{cd07}, and references therein],
$$
{\mathcal E}_+ ={\mathcal E}_-; ~~~~
{\dot{M}}_{+} ={\dot {M}}_{-}-{\dot {M}}_{\rm out}
={\dot {M}}_{-}(1-R_{\dot m}); ~~~~
{\Pi}_+={\Pi}_-,
\eqno{(3)}
$$
Assuming the jet to be launched with the same specific energy,
angular momentum and density as the post-shock disc,
the expression for relative mass outflow rate
is given by \citep{cd07},
$$
R_{\dot m}={\dot {M}}_{\rm out}/{\dot {M}}_{-}
=\frac{Rv_{j}(x_s) {\mathcal A}(x_s)}
{4 \pi \sqrt{\frac{2}{\gamma}}x^{3/2}_s (x_s-1)a_+u_{-}},
$$
where, the compression ratio is defined as $R={\Sigma_+}/{\Sigma_-}$.
Since, the information of $R_{\dot m}$ is in the shock condition
itself, we need to solve accretion-ejection equations
simultaneously. The method to do so is as follows: \\
(a) we assume $R_{\dot m}=0, ({\dot {m}_{-}}= {\dot {m}_{+}})$,
and with the supplied values of
($x_{ci}$, $\lambda_{ci}$, ${\alpha}$, ${\beta}$)
we integrate Eqs. (1a-d) outwards along the sub-sonic
branch of the post-shock region. Equation (3) 
is used to compute the pre-shock flow quantities, which are employed
to integrate outwards to find the location of $x_{co}$. 
The location of the jump for which $x_{co}$ exists is the virtual shock
location ($x^{\prime}_s$). \\
(b) Once $x^{\prime}_s$ is found out, we assign ${\mathcal E}_j={\mathcal E}
(x^{\prime}_s)$
and $\lambda_j=\lambda(x^{\prime}_s)$ to solve the jet equations
and compute the corresponding $R_{\dot m}$. \\
(c) We use this value of $R_{\dot m}$ in Eq. (3) and again calculate
the shock location. \\
(d) When the shock locations converge we have the actual shock
location ($x_s$), and the corresponding $R_{\dot m}$ is the
mass outflow rate. \\
In other words, we are launching jets with same ${\mathcal E}$,
$\lambda$, and ${\rho}$ as that of the shock.

Presently, we consider viscosity and synchrotron cooling process as the
source of dissipation in the flow. Viscosity reduces the
angular momentum, while increases the energy as the flow accretes towards
the central object. Cooling process on the other hand, decreases
the flow energy inwards while leaving the angular momentum distribution
unaffected.
For proper understanding of the effect of viscosity and cooling
on determining mass outflow rates we need to fix $({\mathcal E},\lambda)$ 
at some length-scale (around inner or outer boundary),
and then vary ${\alpha}$ and $\beta$.

\begin{figure}

\begin{center}
\includegraphics[width=0.55\textwidth]{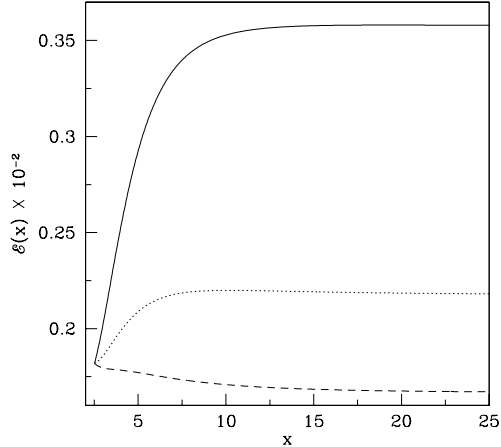}
\end{center}

\caption{ ${\mathcal E}(x)$ with $x$ is plotted for
$\beta=0$ (dashed), $0.01$ (dotted) and $0.036$ (solid).
Other parameters are
(${\mathcal E}_{ci},\lambda_{ci}$)=($0.00182,1.73$) and
$\alpha=0.001$. }
\label{fig2}
\end{figure}

As $x_{ci}$  is very close to the horizon, fixing
(${\mathcal E}_{ci},\lambda_{ci}$) at $x_{ci}$ is almost equivalent
to fixing the inner boundary flow quantities.
In Fig. 2, we plot ${\mathcal E}(x)$ with $x$ for
$\beta=0$ (dashed), $0.001$ (dotted) and $0.0036$ (solid),
where the inner boundary flow
quantities are (${\mathcal E}_{ci},\lambda_{ci}$)=($0.00182,1.73$) and
$\alpha=0.001$.
For the cooling free solution (dashed), the energy of the flow increases
inwards due to viscosity. For solutions with significant cooling
(dotted, solid), the increase in energy due to viscous heating
is completely over shadowed, causing the energy to decrease towards
the black hole.
Increase in cooling efficiency signifies, matter with higher energies
at the outer boundary, falls into the black hole with identical
${\mathcal E}_{ci}$.
If standing shocks form, then under these circumstances
energy at the shock will increase with $\beta$.
In the following, we discuss the role of viscous heating and
synchrotron cooling in determining the mass outflow rate.

\begin{figure}
\begin{center}
\includegraphics[scale=0.5]{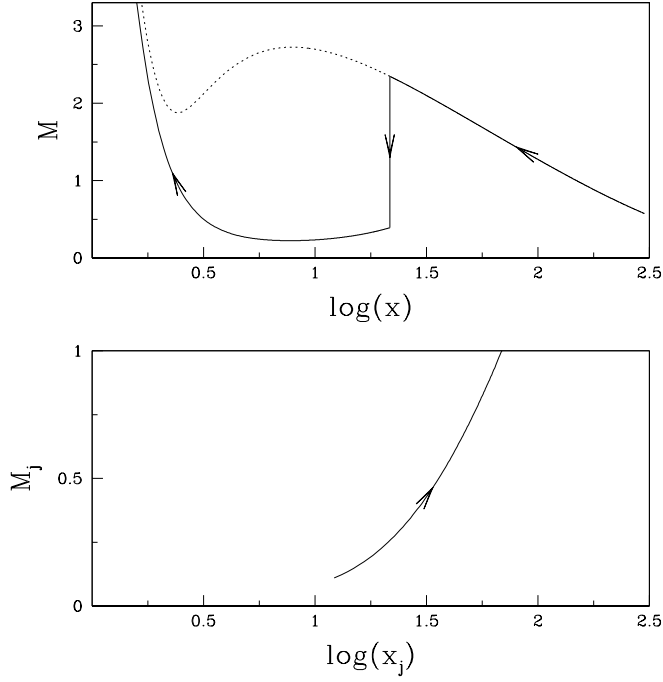}
\end{center}
\caption{Upper panel: Inflow Mach number ($M=u/a$) with $log(x)$.
The inflow parameters are
$x_{ci}=2.444$, $\lambda_i=1.75$,
$\alpha=0.005$, and $\beta=0.01$ where, $x_s=21.64$, ${\mathcal E}_s=0.00175$,
$\lambda_s=1.766$ $x_{co}=166.57$, $\lambda_o=1.799$. The dotted curve
is the shock free solution.
Lower panel: Outflow Mach number
($M_j=v_j/a_j$) with $log(x_j)$,
the outflow critical point $x_{jc}=68.63$ ($r_{jc}=
270.8$), and the jet coordinates at the base
is given by $x_{jb}=12.2$ ($r_{jb}=21.24$).
The relative mass loss rate is $R_{\dot m}=0.0816$.
}
\label{fig3}
\end{figure}

In Fig. 3, we present a global inflow-outflow solution. In
the top panel, the Mach number $M$ of the accretion flow is
plotted with $log(x)$. The solid curve represents shock induced
accretion solution. The inflow parameters are $x_{ci}= 2.444$,
$\lambda_i=1.75$, $\alpha=0.005$, and $\beta=0.01$ (for these
parameters ${\mathcal E}_{ci}=0.0018$). In the lower panel,
the outflow Mach number $M_j$ is plotted with $log(x_j)$.
In presence of mass loss, the shock forms at $x_s=21.64$
denoted by the vertical line in the top panel, and the
outflow is launched with energy and angular momentum at
the shock (${\mathcal E}_s, \lambda_s =0.00175,1.766$).
The outflow is plotted up to its sonic point ($x_{jc}=68.83$),
and the corresponding relative mass outflow rate is
$R_{\dot m}=0.0816$.

\begin{figure}

\begin{center}
\includegraphics[width=0.55\textwidth]{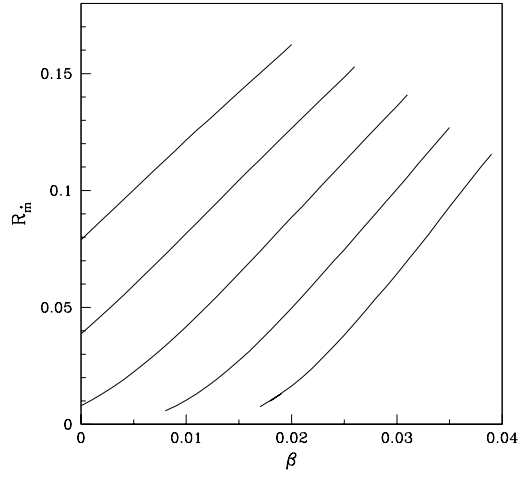}
\end{center}

\caption{ Variation of $R_{\dot m}$ with $\beta$
for $\alpha=0$ --- $0.02$ (left to right with $d{\alpha}=0.005$).
${\mathcal E}_{ci}=0.0018$ and ${\lambda}_{ci}=1.75$.}
\label{fig4}
\end{figure}

To present the global solution, Fig. 3 was obtained only for a set
of input parameters, namely (${\mathcal E}_{ci}$, ${\lambda}_{ci}$,
$\alpha$, $\beta$). We would now proceed to find the explicit
dependence of $R_{\dot m}$ on these parameters. In Fig. 4, we plot
the mass outflow rates ($R_{\dot m}$) with the cooling parameter
${\beta}$, for $\alpha=0$ --- $0.02$ (left to right for 
$d{\alpha}=0.005$). All the curves are drawn for 
${\mathcal E}_{ci}=0.0018$ and ${\lambda}_{ci}=1.75$.
Figure 4 confirms our earlier investigation that $R_{\dot m}$
decreases with increasing viscosity parameter \citep{cd07}.
However, it may be noticed that for fixed $\alpha$,
$R_{\dot m}$ increases with $\beta$. For a given $\alpha$,
the energy at the shock increases with $\beta$ (\eg Fig. 2),
and since the post-shock region (\ie CENBOL) is the base of
the jet, the jets are launched with higher driving force.
This causes $R_{\dot m}$ to increase with $\beta$.
It is to be noted, the two extreme curves (\ie for $\alpha=0.015, 0.02$)
on the right show that, for $\beta=0$ there is no outflow,
but in presence of sufficient cooling steady jets reappear.
As $\alpha$ is increased, $R_{\dot m}$ decreases due to the
gradual reduction of sufficient driving at the jet base,
and beyond a critical $\alpha$ (say, $\alpha_{\rm cri}$)
outflow rate vanishes \citep{cd07}. For flows with
$\alpha>\alpha_{\rm cri}$, the required jet driving 
could be generated by considering sufficiently high $\beta$.
In other words, to get steady outflows in the
realm $\alpha>\alpha_{\rm cri}$, there is a non-zero minimum
value of $\beta$ (say, $\beta_m$) corresponding to each $\alpha$.
Furthermore, for each ${\alpha}$ there is a cut-off in $R_{\dot m}$ at
the higher end of $\beta$ (say, $\beta_{\rm cri}$),
since standing shock conditions are not satisfied there.
Non-steady shocks may still form in those regions, and the investigation
of such phenomena will be reported elsewhere.

\begin{figure}

\begin{center}
\includegraphics[width=0.5\textwidth]{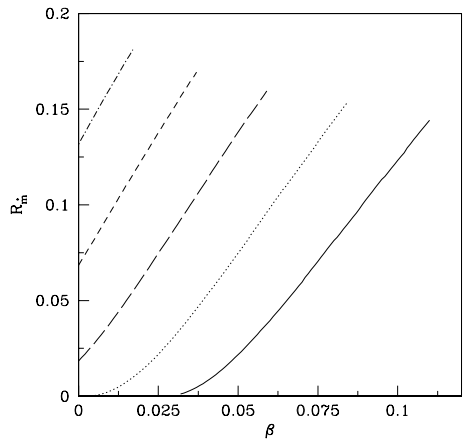}
\end{center}

\caption{ $R_{\dot m}$ is plotted with $\beta$ for
${\mathcal E}_{ci}=-0.001{\rightarrow}0.003$ (right to left, $d{\mathcal
E}_{ci}=0.001$).
Other parameters are $\lambda_{ci}=1.73$ and $\alpha=0.001$.}
\label{fig5}
\end{figure}

In Fig. 5, $R_{\dot m}$ is plotted with $\beta$ for
${\mathcal E}_{ci}=-0.001$ (solid), $0.0$ (dotted), $0.001$ (big dashed),
$0.002$ (small dashed) and $0.003$ (dash-dotted).
Other parameters are $\lambda_{ci}=1.73$ and $\alpha=0.001$.
For a given $\beta$, mass outflow rate increases with ${\mathcal E}_{ci}$.
Higher ${\mathcal E}_{ci}$ corresponds to more energetic flow,
and if these flows produce shock, we get higher $R_{\dot m}$.
On the other hand, even for same ${\mathcal E}_{ci}$,
higher shock energy is ensured with the increase of
$\beta$, and consequently higher $R_{\dot m}$ are produced. 
The solutions corresponding to ${\mathcal E}_{ci}=0$
(dotted) and ${\mathcal E}_{ci}=-0.001$ (solid)
show that $R_{\dot m}{\rightarrow}0$ as $\beta{\rightarrow}0$.
In other words, in presence of cooling, flows with bound energies
at $x_{ci}$ may also produce outflows.
Thus it is clear that shock energy plays an important role in
determining the rate of mass loss from the disc.
Previous studies
of computation of mass outflow rates from inviscid and viscous disc
showed that the angular momentum at the shock dictates the mass outflow rates,
because higher angular momentum produces higher centrifugal driving for
the jet.
This lead us to investigate the role of angular momentum of the
disc in determining the mass outflow rates, when cooling is present. 

\begin{figure}

\hskip -0.5cm\includegraphics[width=0.55\textwidth]{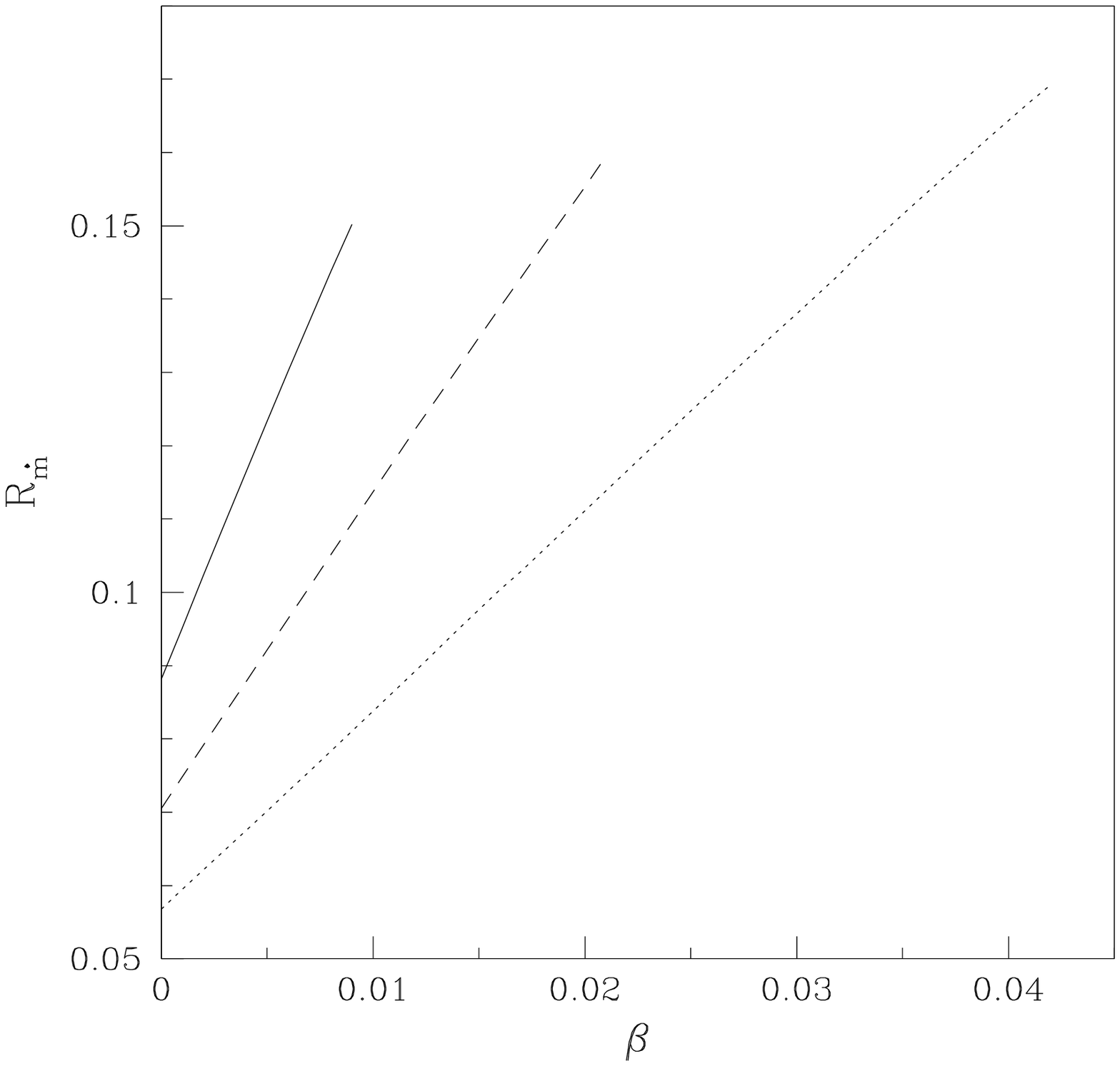}

\vskip -7.86cm
\hskip 7.3cm \includegraphics[width=0.55\textwidth]{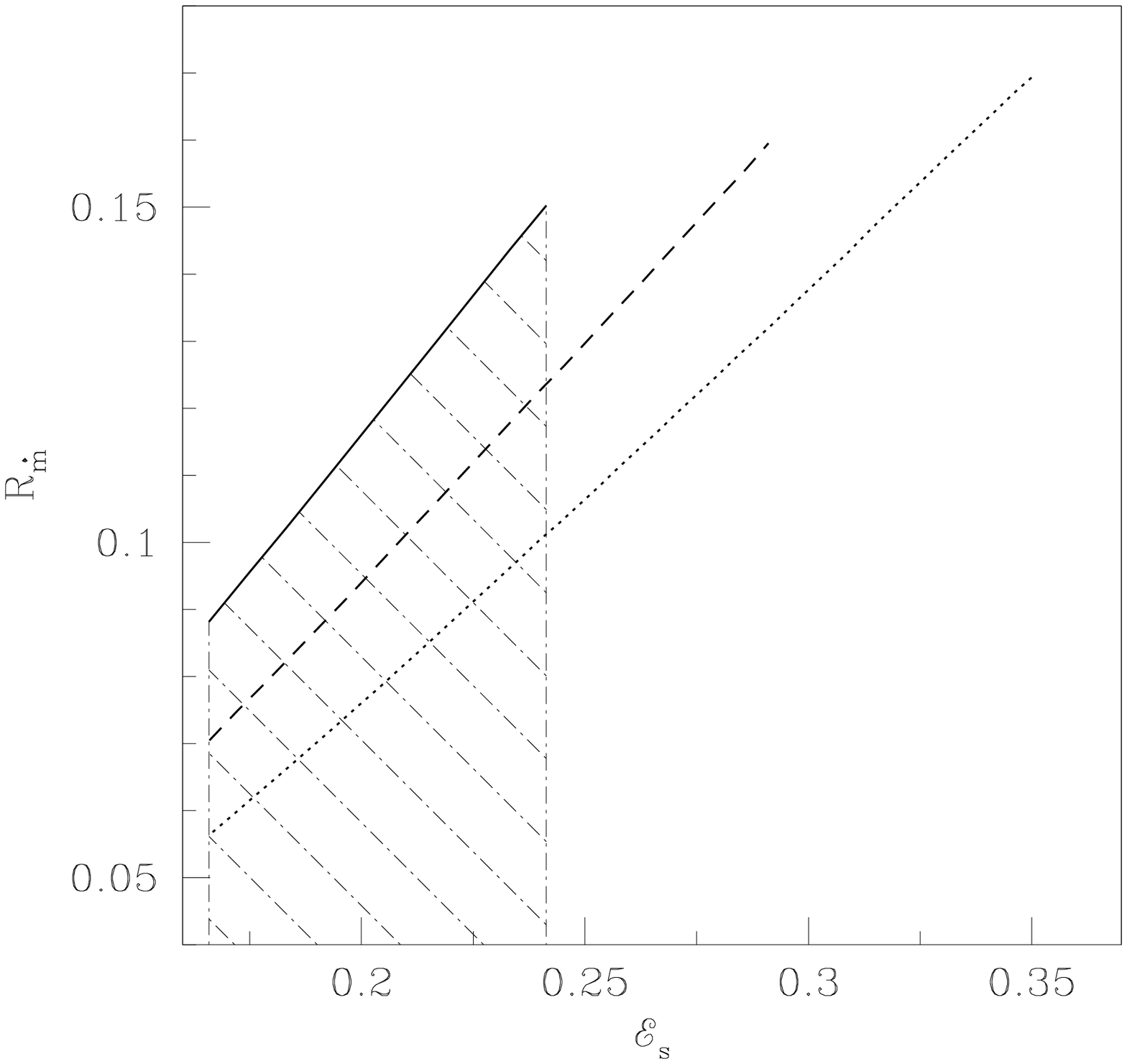}

\vskip 1.0cm
\caption{ (a) Variation of $R_{\dot m}$ with $\beta$
for $\lambda_{ci}=1.73$ (dotted) $1.75$ (dashed) and $1.77$ (solid).
${\mathcal E}_{ci}=0.0018$ and ${\alpha}=0.001$.
(b) Variation of $R_{\dot m}$ with ${\mathcal E}_s$, for parameters
same as Fig. 6a.}
\label{fig6}
\end{figure}

In Fig. 6a, $R_{\dot m}$ is plotted with $\beta$ for
$\lambda_{ci}=1.73$ (dotted),
$1.75$ (dashed) and $1.77$ (solid), where ${\mathcal E}_{ci}=0.00182$,
and $\alpha=0.001$ are kept fixed for all the curves.
For negligible cooling ($\beta \sim 0$), higher angular momentum
flow generates higher $R_{\dot m}$. 
As the centrifugal pressure produces the shock, which in turn drives the jet,
it is not surprising that flows with larger angular momentum will
produce higher $R_{\dot m}$. Similar trend is maintained for nonzero $\beta$.
For a given $\lambda_{ci}$, the energy at the shock (${\mathcal E}_s$)
increases with $\beta$. Thus the combined effects of centrifugal and thermal
driving increase the mass outflow rate.
We do see that there is a cut-off in $R_{\dot m}$ corresponding to
each angular momentum at $\beta{\geq}\beta_{\rm cri}$.
For lower angular momentum flow $\beta_{\rm cri}$ is higher.
To illustrate the effects of thermal driving and centrifugal driving of the jet,
in Fig. 6b, we have plotted $R_{\dot m}$ with ${\mathcal E}_s$
for $\lambda_{ci}=1.73$ (dotted),
$1.75$ (dashed) and $1.77$ (solid), for the same set of
${\mathcal E}_{ci}$ and $\alpha$ as in the previous figure.
It is to be remembered that ${\mathcal E}_s$ is not a new parameter but
is calculated at the shock for the same range of $\beta$ variation
as in Fig. 6a.
In the shaded region, $R_{\dot m}$ is higher for
higher $\lambda_{ci}$. As long as the shock energy
is similar, higher angular momentum results in greater centrifugal
driving for the outflowing matter.
However, lower angular momentum flow can sustain
higher energies across the shock [\eg Fig. 3 of \citet{dcc01}].
For high enough ${\mathcal E}_s$, the thermal
driving starts to dominate over the centrifugal pressure,
and results in higher $R_{\dot m}$ even for lower angular momentum
flow.

\begin{figure}

\begin{center}
\includegraphics[width=0.5\textwidth]{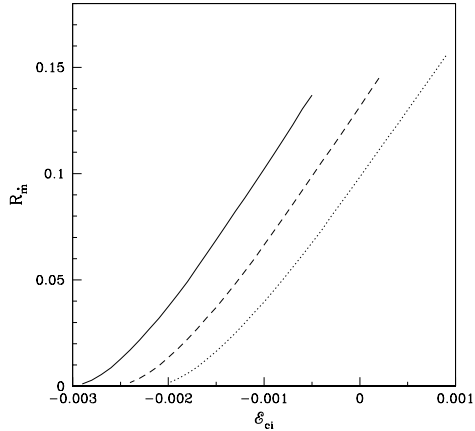}
\end{center}

\caption{ $R_{\dot m}$ is plotted with ${\mathcal E}_{ci}$ for
$\lambda_{ci}=1.73$ (dotted), $\lambda_{ci}=1.74$ (dashed),
$\lambda_{ci}=1.75$ (solid).
Other parameters are $\alpha=0.001$, and $\beta=0.06$.}
\label{fig7}
\end{figure}

In Fig. 7, $R_{\dot m}$ is plotted as a function of ${\mathcal E}_{ci}$,
for various values of $\lambda_{ci}=1.73$ (dotted),
$\lambda_{ci}=1.74$ (dashed), $\lambda_{ci}=1.75$ (solid). 
The other flow parameters are $\alpha=0.001$ and $\beta=0.06$.
This figure distinctly shows that even if the accreting flow
starts with unbound energy and produces shock induced outflow,
significant cooling closer to the black hole turns the unbound
energy to bound energy. 

\section{Astrophysical application}

In our solution procedure, we have employed
three different constant parameters $\beta$, $\eta$ and ${\dot m}$
to determine the cooling process.
A cooling mechanism might depend on various other physical
processes apart from
its usual dependence on the flow variables.
In general, ${\dot m}$ regulates cooling, however
to obtain a cooling free solution one needs to consider ${\dot m}=0$, which
is meaningless. We have simplified all such complications by introducing
$\beta$ as a control-parameter for cooling.
A simple inspection of Eq. (1d), shows that for a given
set of ($u$, $a$, $x$), identical cooling rates
may be obtained by rearranging the values of $\beta$, $\eta$ and ${\dot m}$.
It must be noted that, introduction of $\beta$ and $\eta$
do not increase the parameters of our solution, instead these are used
to control the cooling efficiency and the magnetic field strength,
about which there is no prior knowledge. In the previous section,
we have fixed the values of ${\dot m}_{-}$ and ${\eta}$, and controlled
the cooling term by ${\beta}$.
In this section, we have fixed the value of ${\beta}$ to unity, and
allowed physical parameters, such as ${\dot m}_+$ and ${\eta}$ to
dictate the cooling term.

It is a matter of interest to estimate how much matter, energy
and angular momentum enter into the black hole. In the present
paper, the amount of mass fed to the disc is given by 
${\dot m}_{-}$. The rate at which matter is being accreted
into the black hole and the rate of mass loss are self-consistently
computed as ${\dot m}_{+}$ and $({\dot m}_--{\dot m}_+)$. It has
been shown in \citet{cd07} that the specific angular momentum of the
flow close to the horizon, is almost same as ${\lambda}_{ci}$. The actual value of
${\mathcal E}$ close to the black hole should be slightly higher
than ${\mathcal E}_{ci}$. One has to quote the actual value of
${\mathcal E}$ close to the horizon. However, these numbers are
obtained using pseudo-Newtonian potential and may not be
consistent as general relativistic effects are important at such
distances.

We have applied our formalism to calculate the mass outflow rates
from two black hole candidates M87 and Sgr A$^*$.
M87 is supposed to harbour a super massive
black hole [$M_{\rm BH}=3{\times}10^9M_{\odot}$ \citep{fetal94}].
The estimated accretion rate is
${\dot M}_-{\sim}0.13M_{\odot}$yr$^{-1}$ \citep{reynetal96}.
The mass of the central black hole and the accretion rate
of Sgr A$^*$ are $M_{\rm BH}=2.6{\times}10^6M_{\odot}$ \citep{s02}
and 
${\dot M}_-{\sim}8.8{\times}10^{-7}M_{\odot}$yr$^{-1}$ \citep{yuanetal02}.
The accretion disc around the black hole in
Sgr A$^*$ is supposed to be radiatively
in-efficient and of higher viscosity \citep{f99}.
For both the cases we have set $\beta=1$, so the cooling mechanism
is purely dictated by ${\dot m}$ and $\eta$.
To simplify further, we have chosen $\eta=0.01$ for both the
objects.
The accretion rates (in terms of Eddington rate) for M87
is given by ${\dot m}_-=1.89{\times}10^{-2}$ and that for Sgr A$^*$
is ${\dot m}_-=1.47{\times}10^{-4}$, therefore
Sgr A$^*$ is dimmer than M87.  
With proper choice of $\alpha$ and $x_{ci}$,
and $\lambda_{ci}$ (see Table 1),
we compute $R_{\dot m}$ (consequently ${\dot m}_{+}$)
for both the objects mentioned above.
The typical size of such a sub-Keplerian disc should be around a thousand
Schwarzschild radii
across the central object. Accordingly we have set the outer boundary
at $X_T=500r_g$, and have provided the typical value of angular momentum
at such distance ($\lambda_T$) for both the objects.
For M87, the computed values of mass outflow rate and shock location
are $R_{\dot m}=0.073$ and $x_s=40.57$.
In case of Sgr A$^*$, the estimated values of mass outflow rate and
shock location
are $R_{\dot m}=0.1049$ and $x_s=14.415$.

\begin{table}
{{\bf Table 1}: Predicted values of $R_{\dot m}$ and jet power for M87 and Sgr A$^*$.}
\vskip 0.2cm
             \begin{tabular}{|c|c|c|c|c|c|c|c|c|c|c|}
                  \hline
{Object} & ${M}_{\rm BH}$ & ${\dot M}_-$ & $\alpha$&
$x_{ci}$ & $\lambda_{ci}$ & ${\dot m}_+$ & $x_s$ & $\lambda_T$ & $R_{\dot m}$ & $L^{max}_{jet}$ \\
&$M_{\odot}$&$M_{\odot}$/yr&&$r_g$&$cr_g$&${\dot M}_{Edd}$&$r_g$&$cr_g$&\%&erg/s \\
                   \hline
M87 & $3.0$ & $0.13$ & $0.010$ & $2.367$ & $1.78$ & $ 1.75$ & $40.57$ & $2.01$
& $7.3$ & $5.36$ \\ 
& ${\times}10^9$ & & & & & ${\times}10^{-2}$ & & & & ${\times}10^{44}$ \\
                   \hline
Sgr A$^*$ & $2.6$ & $8.80$ & $0.015$ & $2.548$ & $1.71$ & $1.32$ & $14.42$ & $2.44$
& $10.5$ & $5.2$ \\
& ${\times}10^6$ & ${\times}10^{-7}$ & & & & ${\times}10^{-4}$ & & & & ${\times}10^{39}$ \\
                   \hline
\end{tabular}

\end{table}

Assuming the jet's luminosity is significant only at the lobes
(where, the jet energy is mostly dissipated),
the maximum luminosities of M87 and Sgr A$^*$ jets, estimated
from the computed values of respective $R_{\dot m}$, are given in Table 1.
Considering 10\% radiative efficiency at the jet lobe
the jet-luminosities for both M87 and Sgr A$^*$,
agree well with the observed
values \citep{reynetal96,fb99}. Moreover, the size of the computed
jet base for M87 is ${\sim}2x_s{\sim}80r_g$.
\citet{bir99} and \citet{bir02} have estimated the base of jet to be less than
$100r_g$ from the
central black hole, and probably greater than $30r_g$. Evidently our estimate
of the jet base agrees quite well with the observations.
There is no stringent upper limit of the jet base for Sgr A$^*$,
however, our computation gives a result which is acceptable in the
literature \citep{f99}.
We have also provided an estimate of angular momentum at $X_T$. For 
Sgr A$^*$, our estimated $\lambda_T$ is comparable with the result of
\citet{cm97}. However, no reliable estimate of $\lambda_T$ for M87
is currently available.
In terms of physical units, various flow variables for M87 are given by,
${\dot M}_{\rm out}{\sim}0.009M_{\odot}{\rm yr}^{-1}$,
${\dot M}_+{\sim}0.119M_{\odot}{\rm yr}^{-1}$,
${\mathcal E}_{ci}=3.1{\times}10^{17}$erg g$^{-1}$,
$x_s{\sim}3.61{\times}10^{16}$cm,
$\lambda_{ci}{\sim}4.75{\times}10^{25}$cm$^2$s$^{-1}$,
and $\lambda_T{\sim}5.36{\times}10^{25}$cm$^2$s$^{-1}$.
Similarly for Sgr A$^*$,
${\dot M}_{\rm out}{\sim}9.1{\times}10^{-8}M_{\odot}{\rm yr}^{-1}$,
${\dot M}_+{\sim}7.77{\times}10^{-7}M_{\odot}{\rm yr}^{-1}$,
${\mathcal E}_{ci}=4{\times}10^{18}$erg g$^{-1}$, $x_s{\sim}1.11{\times}10^{13}$cm,
$\lambda_{ci}{\sim}3.96{\times}10^{22}$cm$^2$s$^{-1}$,
and $\lambda_T{\sim}5.64{\times}10^{22}$cm$^2$s$^{-1}$.

In this paper, only sub-Keplerian matter distribution is chosen for the
accretion disc. 
However, \citet{ct95} and \citet{cm06} have shown that if a mixture of
Keplerian and
sub-Keplerian matter is chosen, then the spectral properties of the disc
is better understood. These assertions have been ratified for several
black hole candidates \citep{s01a, s01b}.
Since matter close to the black hole must be sub-Keplerian, therefore
regardless of their origin, Keplerian and sub-Keplerian matter mixes
to produce sub-Keplerian flow before falling onto the black hole.
Such transition from two component to single component flow has 
been shown by various authors [\eg Fig. 4b, of \citet{dcnc01}].
The region where such transition occurs may be called `transition
radius' ($X_T$). It must be noted that, $X_T$ is treated 
as the `outer edge' of the disc in our formalism described so far.
The energy (${\mathcal E}_T$) and angular momentum
($\lambda_T$) at $X_T$ can then easily be 
expressed in terms of the accretion rate of the Keplerian component
(${\dot M}_K$)
and the sub-Keplerian component (${\dot M}_{SK}$) \citep{dcnc01}.
Once $X_T$, ${\mathcal E}_T$, $\lambda_T$ is known and the net
accretion rate being ${\dot M}={\dot M}_{SK}+{\dot M}_K$, it is easy to
calculate $R_{\dot m}$ following our formalism.
Thus, it is possible to predict
$R_{\dot m}$ from the spectrum of the accretion disc,
if formalism of \citet{ct95} is applied on our solutions.

\section{Concluding Remarks}

The main goal of this paper was to study how dissipative processes affect
the jet generation in an advective disc model. \citet{cd07} have shown that
mass outflow rates decrease with increasing viscosity parameter. 
In the present paper, we have investigated how the mass outflow
rate responds to the synchrotron cooling. 
The general method of the solution (succinctly described in Section 3.)
is to supply $x_{ci}$, $\lambda_{ci}$, $\alpha$, $\beta$ and then integrate
outwards to find the shock location (and consequently the mass outflow rate).
Needless to say, once the above four parameters are fixed, the solution
determines flow with unique outer boundary (\ie at $X_T$).
Of the four parameters, if $\alpha$ is increased, the solution
corresponds to flow with higher angular momentum and lower energy
at the outer boundary. On the contrary, when $\beta$ is increased then
the solution corresponds to higher energy but identical angular momentum
flow at the outer boundary.
Consequently, more energetic flows
are allowed to pass through standing shock for higher $\beta$, and hence
stronger jets are produced.
We have also shown that, if cooling efficiency is increased, then it is
possible to produce jets even for those $\alpha$-s for which $R_{\dot m}$ is
zero (\eg Fig. 3).
Furthermore, it has been shown that the jets are primarily
centrifugal pressure driven even in presence of cooling.
We notice that standing shocks in higher angular momentum flow
do not exist for higher
cooling efficiency, therefore steady jets are not produced.
However, for higher $\beta$, low angular momentum flow can generate
high enough relative mass outflow rates.

We have applied our formalism on a couple of black hole candidates,
namely, Sgr A$^*$ and M87.
Using the available accretion parameters of the above
two objects as inputs, we have shown 
that one can predict observational estimates of jet power.
Moreover, the typical size of the jet base ($\sim 2x_s$) also agrees
well with observations.
\citet{lb05} had dealt with these two
particular objects, with their methodology which also involve
shocked accretion disc. The methodologies of the present paper and the work of
Le \& Becker
(2005) is quite different in the sense that, Le \& Becker (2005) dealt
with isothermal shock while our model is based on the adiabatic shock
scenario.
In Le \& Becker (2005), the focus was on calculating the number densities
and energy densities around an isothermal shock of an hot tenuous adiabatic
rotating flow, by first order Fermi acceleration process. The energy lost
at the isothermal shock, drives a small fraction of in falling gas to
relativistic energies.
With the given observational estimates of black hole mass, accretion rate etc
of M87 and Sgr A$^*$, they estimated the Lorentz factors of the jet.
We on the other hand, have computed the thermally driven outflows from the
post-shock disc, where the jets are launched with the local values
(specific energy,
angular momentum and density) of the disc fluid at the shock. With input values
of black hole mass, accretion rate, and proper choice of viscosity parameter,
inner sonic point etc we predict the shock location, the mass outflow rate.
We check whether the predicted values are within the accepted limits or not.
We do not estimate the terminal bulk Lorentz factor,
since we believe one has to 
recast the whole framework into the relativistic domain as well as
employ other accelerating processes (\eg magnetic fields etc).
One may wonder at the veracity of the two different processes employed
to explain the observational estimates of jet quantities of
M87 and Sgr A$^*$, in other words, whether the jets are generated by
post-shock thermal driving (we have not investigated magneto-thermal driving
since this is only hydrodynamic investigation), or the jets are launched
by particle acceleration processes. If one can observationally estimate
the rate at which mass being ejected from the accretion disc, probably 
then one can ascertain the dominant effect behind jet generation.
If it can be established that indeed the rate of mass loss is negligible
compared to the accretion rate then probably the formalism of
\citet{lb05} is the more realistic jet generation mechanism. However,
suffice is to say, various numerical simulation results do show
(for non-dissipative as well as dissipative flows) that post-shock
flow thermally drive bipolar outflows, and our effort has been to
investigate how dissipative processes affect the relative mass outflow rates.

In this paper we have only discussed formation of steady jets,
since we have considered only stationary shocks.
\citet{msc96} have shown that, the periodic breathing of the CENBOL
starts
when the post shock in-fall timescale matches
with the Bremsstrahlung cooling timescale.
Presently, we have considered dissipative processes
which are more effective
in determining shock properties compared to Bremsstrahlung.
Therefore, the dissipative processes considered in this
paper, may trigger comparable or
different shock-instabilities in the disc than that has been reported
earlier \citep{msc96}.
Since, the jet formation is primarily controlled by the
properties of the shock, any non-steady behaviour of the shock will leave its
signature on the jet. In particular, a significant oscillation of the shock
(both in terms of the oscillation frequency and its amplitude) may produce
periodic ejections. We are studying dynamical behaviour of the shock in
presence of viscosity and synchrotron cooling
using fully time dependent simulation and results will be reported elsewhere.

\ack

SD was supported by KOSEF through Astrophysical Research Center
for the Structure and Evolution of the Cosmos (ARCSEC),
and 
IC was supported by the KOSEF grant R01-2004-000-10005-0.
The authors thank
U. Mukherjee for suggesting improvements in the manuscript.

%

\end{document}